\documentclass[showpacs,showkeys,preprintnumbers,amsmath,amssymb]{revtex4}
\usepackage{graphicx}
\usepackage{biograph}        
\usepackage{amsthm}

\newtheorem{theorem}{Theorem}[section]
\newtheorem{corollary}[theorem]{Corollary}
\newtheorem{lemma}[theorem]{Lemma}

\begin{document}

\title{Generalization of the event-based Carnevale-Hines integration scheme
for integrate-and-fire models}


\author{Ronald A.J. van Elburg}
	\email{RonaldAJ@vanelburg.eu}
 \affiliation{Department of Artificial Intelligence, Faculty of Mathematics and Natural Sciences, \\ University of Groningen, P.O. Box 72, 9700 AB, Groningen, The Netherlands}
 \affiliation{ Department of Experimental Neurophysiology, Center for Neurogenomics and Cognitive Research,\\ Vrije Universiteit Amsterdam, De Boelelaan 1085, 1081 HV
Amsterdam, The Netherlands 
}
\author{Arjen van Ooyen}%

\affiliation{ Department of Experimental Neurophysiology, Center for Neurogenomics and Cognitive Research,\\ Vrije Universiteit Amsterdam, De Boelelaan 1085, 1081 HV
Amsterdam, The Netherlands 
}

\date{11 May 2008}

\begin{abstract}
\label{abstract}
An event-based integration scheme for an integrate-and-fire neuron model with exponentially decaying excitatory synaptic currents and double exponential inhibitory synaptic currents has recently been introduced by Carnevale and Hines. This integration scheme imposes non-physiological constraints on the time constants of the synaptic currents it attempts to model which hamper the general applicability. This paper addresses this problem in two ways. First, we provide physical arguments to show why these constraints on the time constants can be relaxed. Second, we give a formal proof showing which constraints can be abolished. This proof rests on  a generalization of the Carnevale-Hines lemma, which is a new tool for comparing double exponentials as they naturally occur in many cascaded decay systems including receptor-neurotransmitter dissociation followed by channel closing. We show that this lemma can be generalized and subsequently used for lifting most of the original constraints on the time constants. Thus we show that the Carnevale-Hines integration scheme for the integrate-and-fire model can be employed for simulating a much wider range of neuron and synapse type combinations than is apparent from the original treatment.   

\end{abstract}
\maketitle

\section{Introduction}
\label{sec:intro}
One of the most salient features of neurons is their ability to summate synaptic inputs arriving from other neurons and to respond with the generation of an action potential or spike when the membrane potential reaches a certain threshold value. After its generation, a spike will generally travel down the neurons axon to serve as an input to other cells, including muscles fibers and neurons. In its most basic form, spike generation is captured by the so-called integrate-and-fire model. This model was first conceived a hundred years ago by Lapicque \cite{Lapicque_JPhysPathGen_1907}. Lapicque modeled the subthreshold behavior of the membrane potential as a capacitance in parallel with a resistor based on the electrical properties of the cell membrane. At that time, the spike generating mechanism was not known, and it was therefore only possible to give a phenomenological description of the process. On the basis of electrophysiological experiments, Lapicque assumed that when the membrane potential reached a threshold value, the cell would generate (fire) a spike and subsequently the membrane potential would  be reset to resting level \cite{Abbott_BrResBull_1999,Lapicque_JPhysPathGen_1907}. Integrate-and-fire models are still widely used today, both in simulations and for the analytical study of neural network dynamics.

The integration scheme we analyze here was introduced by Carnevale and Hines \cite{Hines2002,ISI:000222245900161} for the widely used NEURON simulation environment \cite{Carnevale2006} and is event-based. In  event-based models the synaptic coupling between neurons is mediated by events. Events are triggered by threshold crossings of the integrate-and-fire neurons and subsequently communicated to the postsynaptic cells.  Postsynaptically these events initiate a change in the synapse, which in the most common case lead to excitation or inhibition of the postsynaptic cell. Support for event-based integration methods is available in several other scientific neural simulators ( e.g. NEST, XPP and Mvaspike) \cite{ISI:000250064400006} which makes these simulators possible candidates for the implementation of the scheme discussed here. Besides the Carnevale-Hines scheme many other integration schemes are in use to simulate integrate-and-fire models. These schemes range from purely numerical integration schemes, such as Euler and Runge-Kutta, to numerically exact calculations based on root-finding algorithms for determining when the membrane potential crosses the spiking threshold \cite{ISI:000238258400010,ISI:000071831400011,ISI:000242803200002,ISI:000239341000006,ISI:000237873900029,ISI:A1993LU28000042}.   The Carnevale-Hines scheme takes a middle ground between the two above mentioned extremes; it uses explicit knowledge of the exact solution to determine whether threshold crossing will occur, but avoids the expensive explicit calculation of the threshold passage time. Instead, the model employs the computationally cheaper Newton iteration to obtain a spike time estimate. For situations where presynaptic cells become active between two firing times and the order of firing of the cells is unknown (for example, due to mild external noise), we expect this scheme to be computational efficient. An analysis of computational  efficiency, however, is outside the scope of this paper.

The Carnevale-Hines scheme is developed to solve an integrate-and-fire model which includes excitatory synapse as exponentially decaying currents and inhibitory synapse as currents following a double exponential function. With these currents, it is in principle possible to describe the main class of excitatory synapses (characterized by AMPA-receptors) and the main class of inhibitory synapses (characterized by GABA receptors). As mentioned before the Carnevale-Hines scheme uses Newton iteration  to estimate the threshold crossing times,or phrased differently to find the events. In their proof of the correctness of the Newton iteration estimate Carnevale and Hines used the following constraints on the time constants of the synapses and the membrane time constants: $\tau_{decay,excitatory} < \tau_{rise,inhibitory} < \tau_{decay,inhibitory} < \tau_{membrane}$ \cite{Hines2002,ISI:000222245900161}. These constraints imposed by the integration method lead to the loss of many physiological relevant realizations of the conceptual model. To elucidate the biological relevance of this point we will discuss the physiological parameter range in the next paragraph. 

In cortical areas of mammals, the excitatory  AMPA currents have a fast rise time between 0.1 and 0.8 ms, followed by a fast decay of 1 to 3 ms \cite{ISI:A1997XZ44900004,ISI:000082343500021}. In these areas, the inhibitory GABAergic currents have a rise time between 1 and 2 ms  \cite{ISI:000179553500017} and a decay time varying from about 5 ms to about 30 ms \cite{ISI:000230135500033}. Also the membrane time constants of different cortical neurons vary over a wide range, from close to 5 ms to well over over 40 ms \cite{ISI:000220500900001,ISI:000236152100056,ISI:000088236600731,ISI:000189005800007}.  These physiological data show that the decay times for AMPA synapses are similar or larger than the GABA rise times and furthermore that the GABA decay times can be larger than the membrane time constant. However, as stated above the original treatment of the Carnevale-Hines integration scheme requires that the excitatory decay time should be smaller than the  inhibitory rise time, and that the inhibitory decay time should be smaller than the membrane time constant.
Consequently, the full physiological range of parameters as found in experiment is not accessible to the standard implementation of the model. These limitations became apparent to us during studies into the effects of GABA-receptor maturation on microcircuit processing, on which we will report elsewhere. This has lead us to the reexamination of the Carnevale-Hines  integration scheme that we present here. 

The aim then of our analysis in this paper is to remove the unphysiological constraints on the time constants while keeping the strength of this event-based integration scheme. Our analysis will proceed in four stages. In the first stage we introduce the basics of the model. In the second stage we provide an analysis of the physics of the problem. In the third stage we offer our improved proof of the Carnevale-Hines lemma relaxing some of its original preconditions. In the fourth stage the actual analysis of the integration scheme is carried out using the generalized Carnevale-Hines lemma to abolish the unphysiological constraints on the decay times. 
\section{Model and integration scheme}
\label{sec:mechanism}
In this section we will expand upon the short introduction of the event based integrate-and-fire model provided in the introduction. Subsequently we discuss the exact solution of the models subthreshold dynamics. Both in our treatment of the model and the exact solution we will in essence follow the original treatment \cite{Hines2002,ISI:000222245900161}, however our notation is slightly adjusted to account for the possibility of having synaptic subtypes. 

\subsection{Event-based synaptic current driven integrate-and-fire model}

The model is event based, such models for neural networks are based on the assumption that communication between neurons is completely dependent on action potentials generated in the axo-somatic region and that as a result the time of occurence of the action potential contains all the information a single neuron communicates to the neurons it innervates. In the model used here an event is generated at the moment the neuron passes the firing threshold and  axonal propagation and presynaptic delays are  accounted for by delivering it with an appropriate time delay typically in the order of a few milliseconds at a postsynaptic cell. The synapse in the postsynaptic cell responds then by generating a synaptic current. Physiologically, the time course of synaptic currents is determined by the association rate of the neurotransmitter to the receptor, the dissociation rate of the neurotransmitter from the receptor,  the removal rate of neurotransmitter from the synaptic cleft and the driving synaptic reversal potential \cite{Destexhe1994}.  For current-based integrate-and-fire models, it is further assumed that the synaptic current follows the receptor opening, while the induced change in driving force due to a changing membrane potential is ignored. Here this approach is followed and excitatory synapses and inhibitory synapses are included as membrane potential independent currents  $E_{\nu}$  and  $I_{\mu}$, respectively. The subscripts $\nu$ and $\mu$  in these expressions indicate the different excitatory and  inhibitory synaptic subtypes. Because the currents follow the opening and closing of the receptors  their time course is completely specified by the average time course of the open receptor-neurotransmitter complex states. To describe these we use the variables $e_{\nu}$ and $i_{\mu}$ for the excitatory and inhibitory synapses, respectively.  On the arrival of an event at an excitatory synapse the receptors in the excitatory synapse transfer to the open channel state instantaneously , reflecting a fast transient AMPA binding dynamics. In the model  this is reflected by adding the weight of the synapse $w_{e,\nu} \geq 0$ to $e_{\nu}$.  On the arrival of an event at an inhibitory synapse, however, we first get a fast  increase of the amount of receptor-neurotransmitter complex in the closed state followed by a transition of these receptor-neurotransmitter complexes to the open state. This change in amount of receptor-neurotransmitter complex is  modelled by adding the weight of the particular connection $w_{i,\mu} \leq 0$ to an auxiliary variable $j_{\mu}$ describing the closed receptor-neurotransmitter complex state, the sign of the weight is used in this operation to indicate the inhibitory nature of the resulting current. After this event handling the variables $e_{\nu}$,$j_{\mu}$ and $i_{\mu}$ evolve according to the following differential equations  linking the slowest relevant time scales of the kinetic models underlying receptor opening and closing of the AMPA and GABA receptor to a time development model for the synaptic currents:
\begin{eqnarray}
	\frac{de_\nu}{dt}&=&-\frac{1}{\tau_{e_{\nu}}}e_\nu\nonumber\\
	\frac{dj_\mu}{dt}&=&-\frac{1}{\tau_{i_{\mu}}}j_\mu \nonumber\\
	\frac{di_\mu}{dt}&=&-\frac{1}{\tau_{\mu}}i_\mu + a_{j_{\mu}}j_{\mu}\nonumber\\
	\label{eq:synaptic_dynamics}
\end{eqnarray}
Next to the already introduced variables $e_{\nu}$,$j_{\mu}$ and $i_{\mu}$ we also see the appearance of the time constants $\tau_{e_{\nu}}$ for the excitatory decay time, and $\tau_{j_{\mu}},\tau_{i_{\mu}}$ for the inhibitory rise and decay time. Next to these time constants which come with  the dynamics we sketched, we see the appearance of a parameter $a_{j_{\mu}}$  where the appearance of $\tau_{j_{\mu}}^{-1}$ might have been anticipated, the parameter $a_{j_{\mu}}$  acts as normalization constant and is chosen such that an event induced change in $j_{\mu}$ by an amount $w_{i,\mu}$ results in a maximal change of $w_{i,\mu}$ in $i_{\mu}$. 
For technical reasons and in line with the names chosen, we will assume that $\tau_{j_{\mu}} < \tau_{i_{\mu}}$, which assumption basically restricts the mechanism underlying the double exponential current to mechanisms in which the slow time constant acts on $i$ and not on $j$. Although this can be at odds with the actual biophysics, it poses no real constraint on this phenomenological model, which in its spike output is only sensitive to the shape of the inhibitory current which is insensitive to an interchange of $\tau_{j_{\mu}},\tau_{i_{\mu}}$ provided the normalization constants are adjusted accordingly . 

In the model, the membrane potential  is represented by the  variable $m$, and the resting potential and spike threshold are identified with $m=0$ and $m=1$, respectively. When the membrane potential deviates from the resting potential, then in the absence of synaptic currents it decays back to resting potential and will do so with the membrane time constant $\tau_{m}$. Furthermore, the excitatory $I_e=\sum_{\nu} a_{e_{\nu}}e_{\nu}$ and inhibitory $I_i=\sum_{\mu} a_{i_{\mu}}i_{\mu}$ synaptic currents  act on the membrane potential. The constants $a_{e_{\nu}}$  and $a_{i_{\mu}}$ are normalization constants and are chosen such that an isolated instantaneous change in $e_\nu$ by an amount $w_{e,\nu}>0$ induces a maximum depolarization of $w_{e,\nu}$ and, similarly, such that an isolated instantaneous change in $j_\nu$ by an amount $w_{i,\nu}<0$ induces a maximum hyperpolarization of $w_{i,\nu}$ \cite{Carnevale2006,ISI:000222245900161}. Alternatively, they can be chosen to normalize the charge transfer \cite{ISI:000071831400011}. Putting everything together, the differential equation describing the membrane potential becomes:
\begin{eqnarray}
	\frac{dm}{dt}&=&-\frac{1}{\tau_{m}}m +\sum_\nu a_{e_{\nu}}e_{\nu}+ \sum_\mu a_{i_{\mu}}i_{\mu}
	\label{eq:vm_decay}
\end{eqnarray}

This  differential equation together with the accompanying differential equations for the synapses has an exact solution for the subthreshold behavior. These differential equations are linear in the currents and therefore the solution is completely analogous with the case for a single excitatory and a single inhibitory current \cite{Carnevale2006,ISI:000222245900161}. Using the reciprocals of the  time constants   $k_{x}=1/\tau_{x}$ we can write the exact solution as follows:
\begin{eqnarray}
e_{\nu}(t)&=&e_{\nu,0}e^{-k_{e_{\nu}} (t-t_0)},\\
j_{\mu}(t)&=&j_{\mu,0}e^{-k_{j_\mu} (t-t_0)},\\
i_{\mu}(t)&=&i_{\mu,0}e^{-k_{i_\mu} (t-t_0)},\nonumber\\
&&+j_{\mu,0} b_{i_\mu}(e^{-k_{i_\mu} (t-t_0)}-e^{-k_{j_\mu} (t-t_0)})\\
	m(t)&=& m_0 e^{-k_m (t-t_0)} \nonumber\\
	&+& \sum_\nu e_{\nu,0} b_{e_\nu}(e^{-k_m (t-t_0)}-e^{-k_{e_{\nu}} (t-t_0)})\nonumber\\
	&+&\sum_\mu i_{\mu,0} b_{i_\mu}(e^{-k_m (t-t_0)}-e^{-k_i (t-t_0)})\nonumber\\
	&+& \sum_\mu j_{\mu,0} b_{i_\mu}  b_{j_\mu}[(e^{-k_m (t-t_0)}-e^{-k_{i_\mu} (t-t_0)})\nonumber\\
	&-& \frac{k_{i_\mu}-k_m}{k_{j_\mu}-k_m}(e^{-k_m (t-t_0)}-e^{-k_{j_\mu} (t-t_0)} )]. \nonumber\\
	\label{eq:mexact}
\end{eqnarray}
In this expression $t_0$ refers to the last time preceding $t$ at which $m$, $e_{\nu}$, $j_{\mu}$ and $i_{\mu}$ were evaluated, the values of those variables at $t_0$ are denoted by $m_0 ,e_{\nu,0},j_{\mu,0},i_{\mu,0}$. The constants $a_{e_{\nu}}$, $a_{j_{\mu}}$  and $a_{i_{\mu}}$ used in the differential equations are absorbed into  new constants $b$ together with part of the $k_x$, i.e. $b_e=a_e/(k_e-k_m), b_i=a_i/(k_i-k_m), b_j=a_j/(k_j-k_i)$. The challenge entailed in this subthreshold behavior is to extract from it the point where the membrane potential crosses threshold, which is the topic of the next section.

\subsection{Carnevale-Hines  integration scheme}
The Carnevale-Hines integration scheme is developed to solve the problem of finding threshold passage times from the exact solution of the model presented in the previous subsection. The scheme cycles through the following steps: in the first step an event arrives at the neuron, in the second step the actual membrane potential and synaptic currents at the arrival time of the event are calculated from the exact solution, in the third step a check takes place on threshold crossing and in the fourth step the actual event is handled by updating synaptic currents and calculating a new threshold crossing estimate based on these currents, in the fifth step a self-event which will arrive at the estimated threshold crossing time is generated. Let us examine steps three and four in slightly more detail. In the third step a test takes place to establish whether the membrane potential  $m$ is close to the threshold $\theta=1$ (i.e. $m>\theta-\epsilon$, with $\epsilon$ an arbitrary number satisfying $0 < \epsilon << \theta$). If sufficiently close to threshold then it is  assumed that the membrane potential reached threshold and the membrane potential $m$ is reset to zero and an event is sent to the synapse of the cells innervated by this cell. In the fourth step the event is handled. If the event was a synaptic activation then the weight of the synapse is, inline with the model description above, added to either an $e_{\nu}$ or an $j_{\mu}$. If the event was a self-event related to an  estimated threshold crossing time then the currents obtained from the exact solution are kept unaltered.  After this update of the synaptic currents an estimate is made about when to evaluate the subthreshold solution again. This estimate is based on  the time derivative of the membrane potential $m'$ which is equal to the total of the synaptic and leak currents. If the current is not depolarizing $m' \leq 0$, then the evaluation of the  membrane potential is postponed until the arrival of a new synaptic event. If the current is  depolarizing $m'>0$, a future spike time  is estimated $t_e$ using Newton iteration  $t_e=(1-m)/m'+t_0$. This estimated time is used in step five to generate another type of event, a self-event that is sent by the neuron to itself and indicates the latest time to which evaluation of the membrane potential for threshold detection can be postponed. For this approach to work it is necessary that the estimated threshold crossing time is before  the actual threshold crossing, so that the exact solution can be used to evaluate the membrane potential at that time and we can simply start our event handling again to detect whether we actually crossed threshold. 
 
The essential assumption of this mechanism is that every Newton iteration step is underestimating the threshold passage time, which allows it to approach the threshold crossing with several iterations without the risk of overestimating it. The discussion in the next section and the subsequent mathematical proof focus on showing that provided the excitatory synaptic currents decay faster than the inhibitory currents ($\tau_{e_\nu} < \tau_{i_\mu}$ for all combinations of $\nu$ and $\mu$) the Newton iteration step is underestimating threshold passage time.

\section{Threshold passage time is underestimated by Newton iteration}
\label{sec:thresholdpassage}

The formal proof we present in this section is largely analogous to the proof in the original analysis \cite{Carnevale2006,Hines2002}, except that we base it on a generalization of the underlying Carnevale-Hines lemma, for which we give a proof here. This generalization allows us to lift the unphysiological constraints in the original analysis.  The actual proof consist of two parts: the first part shows that when at $t_0$ the derivative $m'=dm/dt\leq 0$, the membrane potential will stay below threshold at least until a new event arrives; the second part shows that when at $t_0$ the derivative $m'$ satisfies $m'>0$, the Newton iteration formula $t_e=(1-m)/m'+t_0$ underestimates threshold passage time. Before presenting the formal proof based on the exact solution for $m(t)$, we analyze the physics, which provides better insight into the actual underlying mechanisms.
 
\subsection{When will Newton iteration underestimate threshold passage time? A physical analysis}
\label{sec:IntIBasedArg}
To answer the question when Newton iteration underestimates threshold passage time we need to look at the different components contributing to the membrane potential derivative $m'=dm/dt$. We have three kinds of currents: a leak current, which acts in the direction of the resting membrane potential; excitatory synaptic currents, which depolarize the membrane; and inhibitory currents, which hyperpolarize the membrane. If the Newton iteration estimate of threshold passage time $t_{e}$
\begin{equation}
	t_{e}=(\theta-m_0)/m_0'+t_0 {\ \rm for \ } m_0'>0
	\label{eq:Newton_estimate}
\end{equation}
is required to underestimate threshold passage time, then that puts limitations on the possible time course of the currents contributing to $m'$. If the membrane potential does not cross threshold $\theta$ between $t_0$ and $t_e$, then it satisfies the inequality
\begin{equation}
	m(t)=m_0+\int_{t_0}^{t} m' dt < \theta {\ \rm for \ } t_0< t <t_{e}.
	\label{eq:integrated_current}
\end{equation}
Using simple arguments, we can derive conditions on the time constants for which the sum of synaptic and leak currents is decreasing. From the inequality above we can immediately see that if during  $t_0 < t < t_{e}$ the total current shows no growth, i.e. $m'(t)\leq m_0'$, then no threshold passage will take place in this time interval. Therefore, under the stronger assumption that   the total current is decreasing the Newton iteration will underestimate threshold passage time.

\begin{figure}[ht]
\resizebox{\hsize}{!}{\includegraphics*{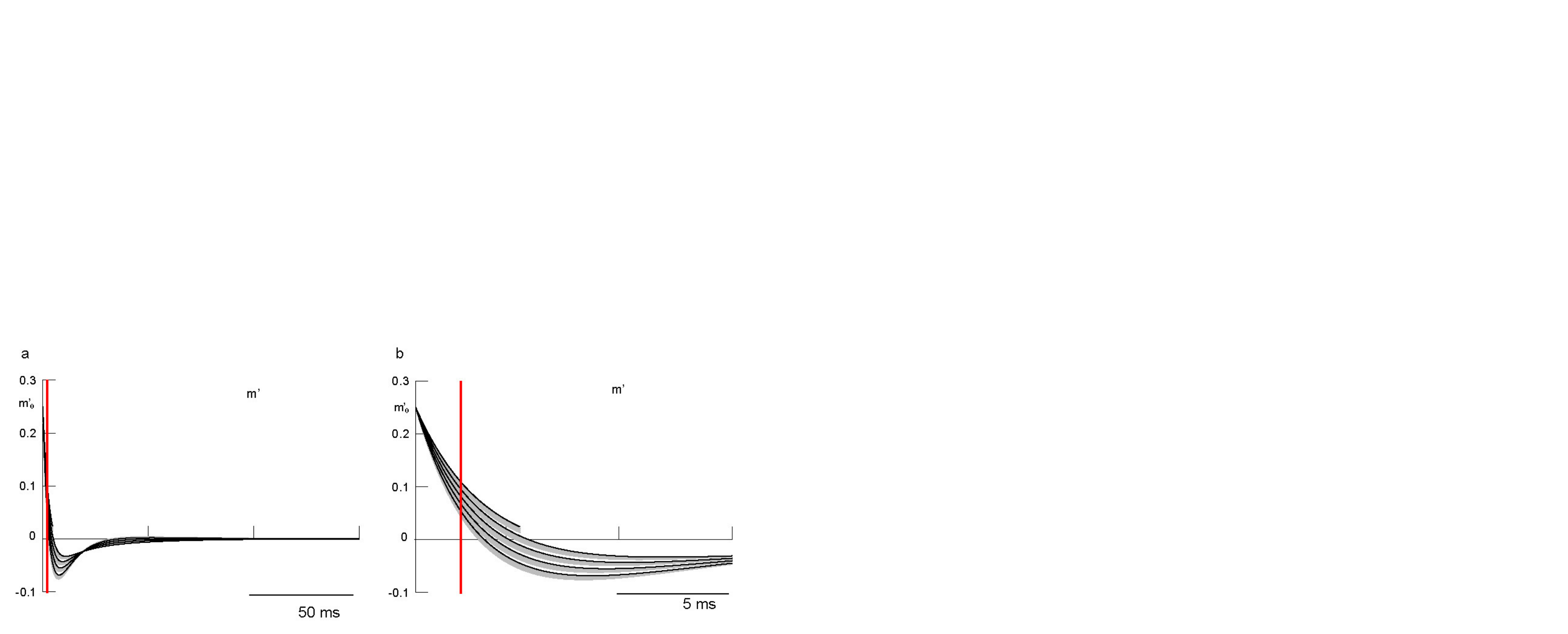}}
\caption{Illustration of the physical analysis. The figures show the derivative of the membrane potential, $m'$, against time. In a., the whole time course is shown, and in b. only the initial part. If $m_0, m_0' > 0$ then $m'$ will stay below $m_0'$ upto the Newton estimated threshold passage time indicated by the vertical red bar. The top black line represents the case were initially all synaptic currents are excitatory and the inhibitory synaptic current is zero for all times; it is shown upto the point were threshold is actually reached. The lower lying black lines are at larger initial inhibitory (and hence excitatory) synaptic currents $(m_{i}'=a_i i=-0.20,-0.15,-0.10,-0.05)$ but without the growth component $(j=0)$. The accompanying grey lines show the effect of adding the growth component $(j=-0.025,-0.050,-0.075,-0.100)$. Parameters used: $\tau_m=30$, $\tau_e=3$, $\tau_i=9$, $\tau_j=2$, $m_0=0.5$, $m_0'=0.25$.
}
\label{fig:integrated_current}
\end{figure}

Let us first analyse the situation in which no inhibitory currents are present (for example the top black lines in figure~\ref{fig:integrated_current}). We know that the excitatory currents are decaying, i.e. their contribution to $m'$ will reduce over time, and the only uncertainty is in the leak currents. If $m'\leq 0$ the excitatory synaptic currents cannot overcome the leak current at the present membrane potential, and therefore after decaying further they are definitely unable to do so.  On the other hand, while $m'>0$ the leak current is growing and therefore reduces $m'$, and we see that $m'<m_0'$ until threshold is reached or until $m'$ reverses sign and the membrane potential moves away from threshold never to reach it. From this we see that either the threshold will never be reached and every finite estimate is underestimating threshold passage time or $m'$ satisfies the inequality $m'(t)\leq m_0'$ upto reaching the threshold and we know then that no threshold passage takes place before the estimated time $t_{e}$.

Now, let us examine the situation in the presence of double exponential inhibitory currents; again examples are  given in figure~\ref{fig:integrated_current}, where grey lines indicate cases where the inhibitory currents are growing and black lines cases where there are only decaying inhibitory currents. The first observation is that a growth of the inhibitory current leads to reduction of $m'$, so adding the growth component associated with $j_{\mu}$ will only strengthen the arguments we can obtain after assuming it to be equal to zero. So assuming $j_{\mu}=0$ and $m_0'>0$ and $m_0 \geq 0$ we know that the individual synaptic currents are decaying towards 0. If, however, the excitatory contribution would decay slowly while the inhibition would decay fast, then the actual resulting synaptic current might be a growing depolarizing current and we might overestimate firing time. It is this consideration that leads to our only real constriction on the time constants: the decay times for excitatory synaptic currents should be faster than those for inhibitory synaptic currents. If, on the other hand, the inhibitory currents decay slower than the excitatory currents, we know that the total synaptic current decreases faster than expected on the basis of the excitatory decay time constants alone and might even become hyperpolarizing.  Again we can see that while $m'>0$ the leak current is growing and therefore also reduces $m'$. Taken together, we find that $m'<m_0'$ until  threshold is reached or until $m'$ reverses sign and the membrane potential moves away from threshold never to reach it. So under the assumption that inhibitory synaptic currents decay slower than excitatory synaptic currents, the condition $m'\leq m_0'$ is fulfilled and we know that the estimated threshold passage time $t_{e}$ based on Newton iteration underestimates threshold passage time. If contrary to our earlier assumption $m_0 < 0$, then we have two separate cases: the first case where the excitatory synaptic current is larger than the inhibitory synaptic current, in which situation all the arguments above apply; and a second case in which we have a dynamics dominated by the leak current. For this leak-dominated phase the argument above does not apply, but because the leak current acts towards the resting potential and not towards the threshold, it is clear that no threshold passage will take place and any estimate will underestimate threshold passage time.  

From these arguments we also see why it is difficult to include NMDA like currents, which are best modeled by an excitatory double exponential current. Their growing excitatory contribution accelerates the rate at which the membrane potential approaches the threshold after activation of the synapse, causing the Newton iteration to overestimate the threshold passage time.
\subsection{Generalized Carnevale-Hines Lemma}
\label{sec:CHLemma}

The exact expression of the membrane potential is built from a large number of double exponential functions. Establishing  upper (lower) bounds on such a sum of double exponentials is strongly simplified by the Carnevale-Hines lemma, which allow us to replace one double exponential with another that for all times is larger (smaller). The lemma follows from the following corollary, which gives us the monotonic development of double exponentials when viewed as a function of one of the decay times: 

\begin{corollary}
\label{corollary1}
For  $t,\mu,\lambda \in \Re$ the functions 
\begin{equation}
	f_\lambda(\mu,t)= \frac{e^{-\lambda t} - e^{-\mu t}}{\mu-\lambda}
	\label{eq:corollary1}
\end{equation}
are defined for $\mu\neq\lambda$ and  by including the limits $\lim_{\mu\rightarrow\lambda} f_\lambda(\mu,t)$ into $f_\lambda$ it can be extended to a function continuous in $\mu$. For fixed $t$ the extended function $f_\lambda$  is a monotonically decreasing continuous function of $\mu$.
\end{corollary}

\begin{proof}
\label{pr:corollary1}
We start with showing that the function $f_\lambda$ can be extended to a continuous function of $\mu$ by adding the point $\mu=\lambda$. The numerator and denominator used in the definition of $f_\lambda$ are $0$ at $\mu=\lambda$, and their derivatives with respect to $\mu$ exists; furthermore, the derivative of the denominator is  equal to $1$ and therefore non-zero. From this we can see that the preconditions of  l'H\^opital's rule are satisfied. From l'H\^opital's rule we know that the limit exist and we find $\lim_{\mu\rightarrow\lambda} f_\lambda(\mu,t)=t e^{-\mu t}$, i.e. the alpha function. After extending $f_\lambda$ with this limit, continuity in $\mu$ follows from observing that l'H\^opital's rule is based on the fact that the left and right limits are equal and hence no discontinuity occurs at $\mu=\lambda$ in the extended $f_\lambda$.

The next step is to prove that the derivative of the function $f_\lambda$ with respect to $\mu$ is negative a.e. for $t \neq 0$. This derivative is given by: 
\begin{equation}
\frac{df_\lambda(\mu,t)}{d\mu}=\frac{e^{-\mu t}}{(\mu-\lambda)^2} \left(t(\mu-\lambda)+1-e^{(\mu-\lambda) t}\right).
\label{eq:dfdmu}
\end{equation}
On the right hand side of this equation the first factor is clearly positive, and therefore we need to prove that the second factor is negative a.e. to show that the function is monotonically decreasing. To show that the second factor is negative a.e., we prove that it has a non-positive maximum at $\lambda=\mu$. If we examine the derivative of the second factor
\begin{equation}
\frac{d((\mu-\lambda)t+1-e^{(\mu-\lambda) t})}{d\mu}=t- te^{(\mu-\lambda) t}
\label{eq:d2ndfac_dmu}
\end{equation} 
we find that it is $0$ for $\mu=\lambda$, positive for $\mu<\lambda$ and negative for $\mu>\lambda$,showing that there is indeed a maximum $(t(\mu-\lambda)+1-e^{(\mu-\lambda) t})=0$ in the second factor at $\lambda=\mu$. As a result we find that ${df_\lambda(\mu,t)}/{d\mu}<0$ for $\mu\neq\lambda$ and the only step left is to prove that  ${df_\lambda(\mu,t)}/{d\mu}$  is continuous.

The expression for ${df_\lambda(\mu,t)}/{d\mu}$ is indeterminate at $\mu=\lambda$, but the preconditions for l'H\^opital's rule are satisfied, indicating that this derivative is continuous at this point. The values of the ${df_\lambda(\mu,t)}/{d\mu}$ at $\mu=\lambda$ are most easily calculated by inserting the Taylor expansion for $t(\mu-\lambda)+1-e^{(\mu-\lambda) t}$ into equation \ref{eq:dfdmu}:
\begin{equation}
\frac{df_\lambda(\mu,t)}{d\mu}=-e^{-\mu t} \sum_{n=2}^\infty (\mu-\lambda)^{n-2} \frac{t^n}{n!} 
\label{eq:limit_dflambdadmu_taylor}
\end{equation}
The indeterminacy at $\mu=\lambda$ is now canceled and we find 
\begin{equation}
lim_{\mu\rightarrow\lambda}\frac{df_\lambda(\mu,t)}{d\mu} = -\frac{t^2}{2}e^{-\mu t}.
\label{eq:limit_dflambdadmu_value}
\end{equation} 
This expression is negative  and thus the derivative ${df_\lambda(\mu,t)}/{d\mu}$ is negative everywhere. 
\end{proof}

\begin{lemma}[Generalized Carnevale-Hines Lemma]
	\label{lem:CHLemma}
	If $\mu_2>\mu_1$ and $\mu_1,\mu_2 \neq \lambda$ then
	\begin{equation}
	  \frac{e^{-\lambda t} - e^{-\mu_1 t}}{\mu_1-\lambda} \geq \frac{e^{-\lambda t} - e^{-\mu_2 t}}{\mu_2-\lambda}
	  \label{eq:CHLemma}
	\end{equation}
\end{lemma}
\begin{proof}
Equality follows from choosing $t=0$ for which we have $f_\lambda(\mu,0)=0$ for all values of $\mu$.  For $t \neq 0$ the inequality is a direct consequence of the corollary. 
\end{proof} 

\subsection{Movement away from threshold implies threshold will never be reached}
\label{sec:away_from_threshold}
We start this analysis from the exact solution of the membrane potential $m(t)$ given in equation \ref{eq:mexact}. The purpose is to show that the exact solution has an upperbound given by the line $m_e(t)=m_0+m_0'(t-t_0)$. Instead of showing that this is true because $m'(t)\leq m_0'$  as in the physical analysis, we will now show it on the basis of the exact solution itself. 

As before, the terms describing growth of the inhibition, i.e. those terms related to $j_{\mu}$,  can be discarded, but now we use the Carnevale-Hines lemma to achieve this. From the lemma we find that for  $\tau_{j_{\mu}} < \tau_{i_{\mu}}$ the factors,
\begin{equation}
	e^{-k_m (t-t_0)}-e^{-k_{i_\mu} (t-t_0)}- \frac{k_{i_\mu}-k_m}{k_{j_\mu}-k_m}\left(e^{-k_m (t-t_0)}-e^{-k_{j_\mu} (t-t_0)}\right) > 0 
	\label{eq:elim_jmu}
\end{equation}
  found in the exact solution for $m(t)$ (equation:\ref{eq:mexact}) are all positive. The factors multiplying these expression, $b_{i_\mu}$ and $b_{j_\mu}$, are also positive, but the factors $j_{\mu,0}$ are negative. The terms, therefore, in which these appear are negative and we obtain an upperbound for $m$ by dropping these terms:
\begin{eqnarray}
	m(t)& \leq & m_0 e^{-k_m (t-t_0)} \nonumber\\
	&&+ \sum_\nu e_{\nu,0} b_{e_\nu}\left(e^{-k_m (t-t_0)}-e^{-k_{e_{\nu}} (t-t_0)}\right)\nonumber\\
	&&+ \sum_\mu i_{\mu,0} b_{i_\mu}\left(e^{-k_m (t-t_0)}-e^{-k_{i_\mu} (t-t_0)}.\right)\nonumber\\
	\label{eq:mupper1}
\end{eqnarray}
In the next step we use the assumption that $\tau_{e_{\nu}}\leq\tau_{i_{\mu}}$ for all combinations of $\mu$ and $\nu$ to be able to apply the Carnevale-Hines lemma. Using an arbitrary excitatory decay time $\tau_{e_{\nu'}}$ we replace the dependence on $\tau_{i_{\mu}}$  in the $i_{\mu,0}$  related terms with a $\tau_{e_{\nu'}}$ dependence. If we fix $\tau_{e_{\nu'}}$ by choosing the largest excitatory decay time for it, we can use the lemma a second time and replace the $\tau_{e_{\nu}}$ dependence in the  $e_{\nu,0}$ related terms  with a $\tau_{e_{\nu'}}$ dependence as well: 
\begin{eqnarray}
	m(t)& \leq & m_0 e^{-k_m (t-t_0)} \nonumber\\
	&&+ \left(\sum_\nu e_{\nu,0} a_{e_\nu}+ \sum_\mu i_{\mu,0} a_{i_\mu}\right)\frac{e^{-k_m (t-t_0)}-e^{-k_{e_{\nu'}} (t-t_0)}}{k_{e_\nu'}-k_m}
	\label{eq:mupper2}
\end{eqnarray}
Now we can use $m'=-k_{m}m +\sum_\nu a_{e_{\nu}}e_{\nu}+ \sum_\mu a_{i_{\mu}}i_{\mu}$ to obtain:
\begin{eqnarray}
	m(t)& \leq & m_0 e^{-k_m (t-t_0)} \nonumber\\
	&&+ m_0 \frac{k_m}{k_{e_\nu'}-k_m}(e^{-k_m (t-t_0)}-e^{-k_{e_{\nu'}} (t-t_0)})\nonumber\\
	&&+ m' \frac{1}{k_{e_\nu'}-k_m}(e^{-k_m (t-t_0)}-e^{-k_{e_{\nu'}} (t-t_0)})
	\label{eq:mupper3}
\end{eqnarray}
We assumed that $m'\leq 0 $, and because the other factors multiplying it are all positive, we can remove the associated  term from the inequality, so that we obtain,
\begin{eqnarray}
	m(t)& \leq & m_0 e^{-k_m (t-t_0)}\nonumber\\
	&&+ m_0 \frac{k_m}{k_{e_\nu'}-k_m}(e^{-k_m (t-t_0)}-e^{-k_{e_{\nu'}} (t-t_0)})
	\label{eq:mupper4}
\end{eqnarray}
This expression is now fully equivalent to the one found in the original treatment \cite{Carnevale2006,ISI:000222245900161}, where it is shown that the derivative with respect to time of the term multiplying $m_0$ is negative; and because it is $1$ at $t=t_0$, it will be smaller than $1$ at larger times while it is also positive. Because $m_0<1$ this shows that $m(t)$ stays between $-\infty$ and $1$ and hence below threshold. 

\subsection{Movement toward threshold shows slowdown}
\label{sec:towards_threshold}
If $m'> 0 $ the step from equation \ref{eq:mupper3} to equation \ref{eq:mupper4} is not allowed, but using our corollary we can replace the last term in equation \ref{eq:mupper3} with the alpha function belonging to the slowest decay time out of $\tau_m$ and $\tau_{e,\nu'}$:
\begin{eqnarray}
	m(t)& \leq & m_0 e^{-k_m (t-t_0)} \nonumber\\
	&&+ m_0 \frac{k_m}{k_{e_\nu'}-k_m}(e^{-k_m (t-t_0)}-e^{-k_{e_{\nu'}} (t-t_0)})\nonumber\\
	&&+ m' (t-t_0) e^{-min(k_m,k_{e_{\nu'}}) (t-t_0)}
	\label{eq:mupper5}
\end{eqnarray}
By the same argument as used before, the first two terms combined are smaller than or  equal to $m_0$; also   $e^{-min(k_m,k_{e_{\nu'}}) (t-t_0)}$ is smaller than $1$, so taking this together with $m_0,(t-t_0) \geq 0$ we obtain:
\begin{equation}
	m(t)\leq m_0 + m' (t-t_0) 
	\label{eq:Newton_it1}
\end{equation} 
When $m_0 < 0$, the second term in the right-hand side of equation \ref{eq:mupper5} is negative and we can drop it from the inequality, which is the correct thing to do if $\tau_{e,\nu'}< \tau_{m}$. If $\tau_{e,\nu'}> \tau_{m}$ we can reorganize the first two terms in such away that we can replace it with  $e^{-k_{e,\nu'} (t-t_0)}$:
\begin{eqnarray}
	m(t)&\leq& m_0 \left(e^{-k_m (t-t_0)} + \frac{k_m}{k_{e_\nu'}-k_m}(e^{-k_m (t-t_0)}-e^{-k_{e_{\nu'}} (t-t_0)})\right)\nonumber\\
		&&+ m' (t-t_0) e^{-min(k_m,k_{e_{\nu'}}) (t-t_0)}\nonumber\\
		&=& m_0 \left(\frac{k_e}{k_{e_\nu'}-k_m}(e^{-k_m (t-t_0)}-e^{-k_{e_{\nu'}} (t-t_0)}) + e^{-k_{e,\nu'} (t-t_0)} \right)\nonumber\\
		&&+ m' (t-t_0) e^{-min(k_m,k_{e_{\nu'}}) (t-t_0)}\nonumber\\
	\label{eq:mupper6}
\end{eqnarray}
From which we see that we have a free choice on whether we want to keep the factor $e^{-k_{e,\nu'} (t-t_0)}$ or $e^{-k_{m} (t-t_0)}$ from the first two terms. It will be convenient to make a choice which fits the last term
\begin{eqnarray}
	m(t)&\leq&  ( m_0 + m' (t-t_0) )e^{-min(k_m,k_{e_{\nu'}}) (t-t_0)}
	\label{eq:mupper7}
\end{eqnarray}
This expression shows that in the area where the linear extrapolation is below zero we have no guarantee that $m(t)$ is under the linear extrapolation, but the actual upper bound is below zero and no threshold passage occurs. When the linear extrapolation crossed zero, we again find $m(t)\leq  m_0 + m' (t-t_0)$, and we see that Newton iteration is underestimating the threshold crossing time. This seemingly strange behavior of the upperbound is not indicating a magical zero crossing of $m(t)$ at the point where $m_0 + m' (t-t_0)$ changes sign, but reflects our lack of knowledge about the system. If the system was dominated by excitatory currents during the whole period, then we know from our physical analysis that the linear extrapolation was above the real curve all the time. If however the system was dominated by an unmasked leak current, then the real curve went above the linear extrapolation, but because the whole movement is towards resting potential and will never cross the resting potential, the linear extrapolation has to cross the membrane potential before passing through zero.  

\subsection{Summary}
\begin{figure}[ht]
\resizebox{\hsize}{!}{\includegraphics*{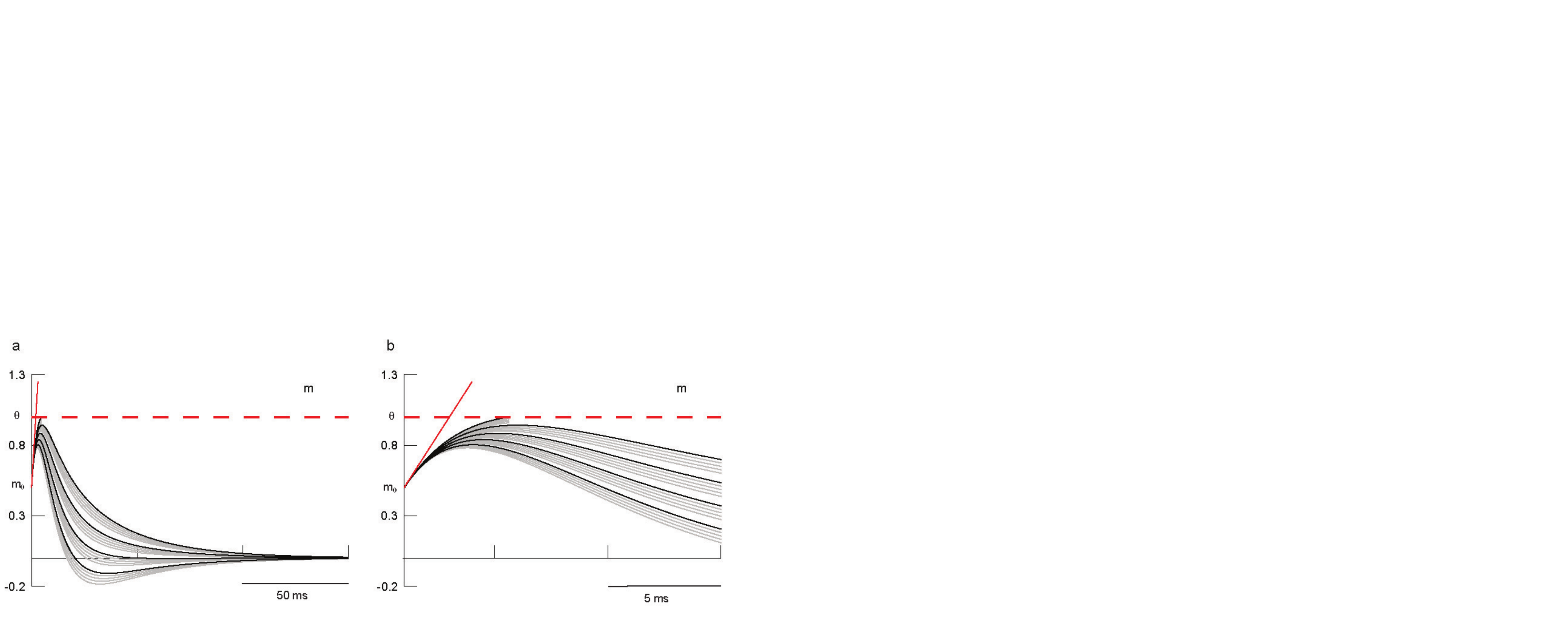}}
\caption{Illustration of the Carnevale-Hines lemma based argument. The figures show the membrane potential, $m$, against time. In a, the whole time course is shown, and in b only the initial part. If $m_0, m_0' > 0$ then $m$ will stay below $m_0+m_0' t$  up to the estimated threshold passage time indicated by the point were the tangent at $t_0$ (red line) crosses threshold $\theta$ (dashed red line). The top black line represents the case were initially all synaptic currents are excitatory and the inhibitory synaptic current is zero for all times; it is shown up to the point were threshold is reached. The lower lying black lines are at larger initial inhibitory synaptic currents $(m_{i}'=a_i i=-0.20,-0.15,-0.10,-0.05)$ but without the growth component $(j=0)$. The accompanying grey lines show the effect of adding the growth component $(j=-0.025,-0.050,-0.075,-0.100)$. Parameters used: $\tau_m=30$, $\tau_e=3$, $\tau_i=9$, $\tau_j=2$, $m_0=0.5$, $m_0'=0.25$.
}
\label{fig:Carnevale_Hines}
\end{figure}

Let us summarize the behavior of the membrane potential in this model as we analyzed it both by the integrated-current based argument and by the Carnavale-Hines lemma. When excitatory synaptic currents dominate (see equations~\ref{eq:Newton_it1},\ref{eq:mupper7}), as in the early stages of the examples given  in figure~\ref{fig:Carnevale_Hines}, the membrane potential increases but always stays below the linear extrapolation . When inhibitory synaptic currents dominate (see equation~\ref{eq:mupper4}), the membrane potential moves away from threshold, and if they are sufficiently strong, as in the lower curves shown in  figure~\ref{fig:Carnevale_Hines}, they lead to reversal of  the polarity of the membrane. When in the last part of these curves leak currents dominate (see equations~\ref{eq:mupper4},\ref{eq:mupper7}), the membrane potential moves  toward resting potential. If the latter takes place at negative membrane potentials, the membrane potential can exceed linear extrapolation, because decay of inhibitory synaptic current can quickly unmask an upward leak current. Although in this situation the linear extrapolation is exceeded, no overestimation of threshold passage time will occur because no threshold passage takes place in the leak dominated phase. In conclusion, we established that either no threshold crossing will take place or that threshold crossing takes place after the Newton estimate for threshold crossing time.

\section{Discussion}
\label{sec:outlook}
We have shown that the non-physiological constraints on the time constants in the event-based integrate-and-fire Carnevale-Hines integration scheme can be relaxed. The Newton iteration step is underestimating threshold passage time provided the slowest decaying excitatory synaptic current $(\nu_s)$ decays faster than the fastest decaying inhibitory current $(\mu_f)$, i.e. we have ($\tau_{e_{\nu_s}} < \tau_{i_{\mu_f}})$ . This makes the model applicable to a much wider range of neuron and synapse combinations in the nervous system. 
Although we here analyzed a current-based model, the argument about threshold passage time, which is based on our physical analysis, can with slight modifications also be applied to conductance-based integrate-and-fire models. Conductance-based models have synaptic reversal potentials, but these do not weaken our argument, as can be seen as follows. On approaching the reversal potential, the driving force for the excitatory current is reduced. The excitatory currents will therefore be reduced further than would be expected from channel kinetics alone. The driving force of the inhibitory currents is enhanced on approaching the threshold potential. The inhibitory currents will therefore be less reduced than would be expected from channel kinetics alone. We see from this that including reversal potentials will only strengthen the threshold passage time argument. The only ingredient missing, then, to apply the Carnevale-Hines integration scheme to conductance-based integrate-and-fire models is the lack of an exact solution for the membrane potential or another numerically efficient way to calculate the membrane potential after a time step of arbitrary size.
Within the Carnevale-Hines integration scheme based on Newton iteration slowly growing excitatory currents (like NMDA currents) cannot be included. We can however, sometimes, replace the Newton iteration by an higher order extrapolation scheme.  Lets for example take a synaptic model for NMDA currents using the same differential equations as are used for the inhibitory current (admittedly ignoring the magnesium block, which is not within reach of our treatment here). We now take $i,j$ to represent the double exponential excitatory currents and auxiliary variable, respectively, and leave out the inhibitory currents. To do this we only need to change the sign of $j$, so we assume $j\geq 0$. If we now further assume that  $\tau_i> \tau_m$ then the exact solution has an upperbound. 
\begin{equation}
m(t) \leq m_0 + m't + j_0 \frac{k_m-k_i}{k_j-k_i}(t-t_0)^2 
\end{equation}
This quadratic upperbound for the exact solution was found using  appropriate $\alpha$-functions as upperbounds for the double exponentials, i.e. by repetitively using the generalized Carnevale-Hines lemma.  We again obtain a threshold crossing estimate if we determine where this upperbound crosses threshold. This new estimate does not solve the problem of unmasking yet, but we expect that unmasking can be  dealt with in a similar way leading to extra terms quadratic terms in the upperboundary. The reason is that only already existing decaying currents can be unmasked, those currents lead to a double exponential contribution to the membrane potential development and putting an upperbound on a combination of two double exponentials involves two times in succession the replacement of a double exponential by an $\alpha$-function leading to a $(t-t_0)^2$ term. If this reasoning is correct then in even more general physiological circumstances, i.e.  $\tau_{decay,AMPA}<\tau_{decay,GABA}<\tau_{decay,NMDA}$ a threshold crossing estimate can be found by taking the positive root of a quadratic polynomial.

\acknowledgements RvE was supported by the   Computational Life Sciences Program of the Netherlands Organization for Scientific Research (NWO, CLS2003, 635.100.000.03N36) and by the Dutch Companion Project funded by the Dutch agency for innovation and sustainable development SenterNovem (SenterNovem, IS053013).

\bibliography{IntFire4}   

\end{document}